\documentstyle[prl,aps,twocolumn,epsf]{revtex}
\newcommand{\lapx}{\mbox{\raisebox{-4pt}{$\,\buildrel<\over\sim\,$}}}

\begin{document}
\draft
\title{Current bistability and  hysteresis
in strongly correlated quantum wires} 
\author{R.~Egger$^{1,2}$,  H.~Grabert$^2$, A. Koutouza$^3$,
  H.~Saleur$^3$, and F. Siano$^3$}
\address{${}^1$Institute for Theoretical Physics, University of
California,
Santa Barbara, CA 93106-4030\\
${}^2$Fakult\"at f\"ur Physik, Albert-Ludwigs-Universit\"at,
D-79104 Freiburg, Germany\\
${}^3$Department of Physics, University of Southern California,
Los Angeles, CA 90089-0484
}
\date{Date: \today}
\maketitle
\begin{abstract}
Nonequilibrium transport properties are determined exactly for an
adiabatically contacted single-channel quantum wire 
containing one impurity. Employing the Luttinger liquid model
with interaction parameter $g$, for very strong interactions,
$g\lapx 0.2$, and sufficiently low temperatures, we find 
an S-shaped current-voltage relation. The unstable branch with 
negative differential conductance gives rise to
current oscillations and hysteretic effects. 
These nonperturbative and nonlinear features appear 
only out of equilibrium.
\end{abstract}

\narrowtext

Transport in 1D conductors is one of the focal points
of condensed matter physics.  At low energy scales, such materials
have been predicted long ago to behave as Luttinger liquids (LL) instead
of Fermi liquids \cite{book}.  Over the past few years,  several
possible experimental realizations of LL behavior have been reported.
In particular, narrow quantum wires (QW) in semiconductor
heterostructures can be operated in the single-channel
limit \cite{tarucha95,yacoby96}.  Other realizations
include quasi-1D materials such as long chain molecules \cite{chain}, carbon
nanotubes \cite{tube}, or edge
states in fractional quantum Hall (FQH) bars \cite{chang}.  Since the
latter are in fact chiral LLs, where right- and left-moving branches
are spatially separated, FQH edge state transport 
\cite{kane92,bcft1} is distinct from the case of a quantum wire.
In this Letter, we
emphasize the important and indeed surprising differences arising for
standard (achiral) LL systems characterized by the
interaction parameter $g<1$. We focus on 
the archetype problem of a spinless single-channel QW containing
backscattering (BS) by one impurity \cite{kane92}.

Our main results are as follows.  The current $I(U,T,g)$ under
an applied voltage bias $U$ at temperature $T$ obeys {\em scaling}
[$I$ depends only on $U$ and $T$ measured in terms of the
impurity scale $T_B$] and a {\em duality relation} connecting the
strong and weak BS limits under the simultaneous exchange
$g\to 1/g$.  Both the scaling function  and the 
duality relation are different
from the FQH case and are determined exactly.   
For very strong interactions, $g\lapx 0.2$, and low temperatures,
the $I-U$ characteristics is {\em multivalued}, containing an unstable
branch of negative differential conductance (NDC).  Once the QW
is embedded in a load circuit, this S-shaped $I(U)$ relation 
can lead to hysteresis, current switching and self-sustained 
current oscillations \cite{scholl}.  Such effects could be observed
in a QW of very low electron density. 

{\sl Coupling of QW to voltage bias.~---}
We focus on a QW adiabatically connected to ideal time-independent
voltage reservoirs held at electro-chemical potentials
$\mu_{1,2}$ \cite{landauer,egger96,egger98}.
The 1D conductor extending from $-L/2<x<L/2$  is
considered to be a LL, with an impurity of BS strength
$\lambda$ sitting at $x=0$.  This defines 
$T_B=c_g\lambda(\lambda/\omega_c)^{g/(1-g)}$,
where $\hbar=k_B=1$, $\omega_c$ is the electronic bandwidth,
and $c_g$ is a numerical prefactor of order unity (its precise
value is of no interest here and given in Ref.\cite{saleurlate}).
The Coulomb interactions take the form 
$H_I = (e/2) \int dx \rho(x) \varphi(x)$,
where $\rho(x)$ is the electron density and
the Poisson equation is replaced by 
\begin{equation}\label{poisson}
e\varphi(x) = u_0 \rho(x) \quad {\rm with} \qquad g=(1+u_0/\pi v_F)^{-1/2}\;,
\end{equation}
with the Fermi velocity $v_F$. 
Here a screening backgate or
other metallic surroundings cause 
short-ranged interactions within the QW. 
Since we are dealing with a strongly correlated non-Fermi liquid system,
the well-known Landauer approach \cite{landauer} does not 
apply.  In the past, external voltage sources were often modeled by
attaching $g=1$ LLs to the ends of the QW \cite{schulz}, but computations 
become exceedingly difficult for $\lambda>0$.
Therefore we employ the  radiative
 boundary conditions of Ref.\cite{egger96} which represent
the natural extension of Landauer's original ideas \cite{landauer}
to a strongly correlated 1D metal. 
Using different arguments, these boundary conditions have
been confirmed and generalized to a.c.~transport
\cite{safinew,buttiker}.

Let us briefly summarize the main ideas \cite{egger96,egger98}. 
Suppose one injects the ``bare" densities $\rho_R^0$ and $\rho_L^0$
at positions $x$ close to the end of the QW. 
The average density $\rho=\rho_R+\rho_L$ [we omit the 
expectation values for brevity] is then self-consistently
determined by 
\begin{equation}
\rho_R(x)+\rho_L(x)  = \rho_R^0+\rho_L^0 - e\varphi(x)/ \pi v_F \;,
\end{equation}
since the band bottom shifts by $e\varphi(x)$.
With Eq.~(\ref{poisson}) we then get  the local relation
$\rho_R+\rho_L= g^2 (\rho_L^0+\rho_R^0)$,
in agreement with the compressibility of a LL, $\kappa=
g^2/\pi v_F$.
Since screening affects only the total charge, see Eq.~(\ref{poisson}),
we also have $\rho_R-\rho_L=\rho^0_R-\rho^0_L$.
Solving these two relations for $\rho^0_{R/L}$  and using
$\rho_R^0(-L/2) = \mu_1/2\pi v_F$ and 
$\rho_L^0(L/2) = \mu_2/2\pi v_F$, we obtain the boundary conditions 
of Refs.\cite{egger96,egger98,safinew},
\begin{equation}\label{bc}
\frac{g^{-2}\pm 1}{2}\rho_R(\mp L/2) + \frac{g^{-2}\mp 1}{2}\rho_L(\mp L/2)
= \pm \frac{U}{4\pi v_F} \;,
\end{equation}
where we put $\mu_1=-\mu_2=U/2$ with the applied voltage $U$ and $e=1$.
We assume full translational invariance such that
the sound velocity $v=v_F/g$ (below
we set $v=1$).  The Sommerfeld-like
boundary conditions (\ref{bc}) are imposed at the left/right end of the
conductor at long times $t$ where the stationary nonequilibrium state
has been reached.  Below we assume that the relevant energy scale
($k_B T$ or $eU$) exceeds $v/L$, and henceforth take $L\to \infty$.

{\sl Exact solution and duality.~---}
Like for the tunneling problem in the FQH effect \cite{bcft1},
it is possible to compute exactly the current out of equilibrium. 
The basic idea is first to
fold the problem onto the boundary sine-Gordon model, and then use 
integrability of the latter \cite{saleurlate}.
 The final equations governing the physics
are quite simple. For clarity, we restrict ourselves
to the case $g=1/p$ with $p$ integer.
 The basic quantities are then pseudo-energies $\epsilon_j(\theta)$
for rapidity $\theta$, obeying a set
of thermodynamic Bethe ansatz (TBA)  integral equations,  
\begin{equation} \label{tba}
\epsilon_{j}(\theta)=T\sum_{k}N_{jk}\int d\theta'{s(\theta-\theta')\over 2\pi
  } \ln \left( 1+e^{
(\epsilon_{k}(\theta')-\mu _{k})/T}\right) \;,
\end{equation}
where $s(\theta )=(p-1)/\cosh [(p-1)\theta]$
and  $N_{jk}$ is
the incidence matrix of the following TBA\ diagram, on which the labels
$j,k$ run:

\bigskip
\noindent
\centerline{\hbox{\rlap{\raise28pt\hbox{$\hskip5.2cm\bigcirc\hskip.25cm +$}}
\rlap{\lower27pt\hbox{$\hskip5.1cm\bigcirc\hskip.3cm -$}}
\rlap{\raise15pt\hbox{$\hskip4.8cm\Big/$}}
\rlap{\lower14pt\hbox{$\hskip4.7cm\Big\backslash$}}
\rlap{\raise15pt\hbox{$1\hskip1cm 2\hskip1.3cm\hskip.4cm p-3$}}
$\bigcirc$------$\bigcirc$------$\bigcirc$------$\bigcirc$------$\bigcirc$\hskip.5cm
$p-2$ }}

\bigskip

\noindent
Here we have defined parameters for the breathers 
$m_j=2\sin {j\pi\over 2(p-1)}$
with $j=1,\ldots, p-2$, and for the kink and antikink,
$m_\pm=1$ \cite{bcft1}.   The latter are the fundamental charged
particles in the integrable description. All the physical particles are coupled by non trivial $S$ matrix elements. However, manipulation of the TBA equations gives rise to the simpler structure of coupling for the pseudoenergies represneted on the foregoing diagram.
The chemical potentials are $\mu_j=0$ for the $p-2$ breathers,
and $\mu_\pm=\mp W/2$ for kink and antikink.
Here $W$ is determined self-consistently, see
below.
 Having found the $\epsilon$'s, the densities of kinks and
antikinks are given by 
 $\sigma_{\pm }=nf_{\pm }$, where the pseudo-energies $\epsilon_{\pm}$ 
are equal,   $2\pi n=d\epsilon_{\pm}/d\theta$, and the filling fractions
read 
\begin{equation}\label{distr}
f_{\pm}(\theta)=1/\{1+\exp[(\epsilon (\theta)\mp W/2)/T]\} \;.
\end{equation}
With these definitions, the final expression for the current reads
\begin{equation}
\label{curr}
I=\int \left| T_{++}\right|^{2}(\sigma_{+}-\sigma_{-})\,d\theta \;,
\end{equation}
where, with the impurity scale $T_B\propto \exp \theta_B$ defined above, 
the tunneling probability is
\begin{equation}
|T_{++}|^2= \left( 1+
  \exp[-2\left(g^{-1}-1\right)(\theta-\theta_B)]\right)^{-1} \;.
\end{equation}
The current is  implicitly a function of $W$, which is
self-consistently determined through
\begin{equation} \label{bdrygene} 
\int \left( \left| T_{++}\right| ^{2}+\frac{1}{g}\left| T_{+-}
\right|^{2}\right) (\sigma _{+}-\sigma _{-})\,d\theta =
\frac{U}{2\pi}\;,
\end{equation}
where $|T_{+-}|^2 = 1-|T_{++}|^2$.

>From these equations, it is now straightforward to deduce 
the following identity giving the parameter $W$ in terms of the
physical voltage and current,
\begin{equation}
 \label{defw}
U=2\pi \left(1-{1\over g}\right)I+W 
\end{equation}
In the sequel, we will also use the quantity 
\begin{equation} \label{defv}
V=U-2\pi I= W-2\pi I/g \;,
\end{equation}
 whose physical meaning is the four-terminal voltage 
across the impurity \cite{egger96}, i.e., the voltage
difference measured by weakly coupled reservoirs 
on either side of the impurity.
We then find that for non-vanishing $T_B$,
 the current interpolates between $I=0$ and (going back to physical
 units) $I=(e^2/h) U$ as the applied
 voltage $U$ is increased. Similarly, the linear conductance
interpolates between the perfectly quantized high-temperature
value $G=e^2/ h$
also found in a clean QW \cite{schulz},
and the vanishing low-temperature ($T\ll T_B)$
conductance predicted in Ref.~\cite{kane92}.
For temperatures above $\omega_c$ or $v_F/r$, where $r$ is the
interaction range, the conductance saturates 
before reaching $e^2/h$ in a real QW.

The above equations can be solved 
in closed form at vanishing temperature. The solution 
for arbitrary $g$ is
expressed in terms of two different  series expansions, depending 
on whether the impurity BS is weak or strong. In the latter case,
 we find
\begin{eqnarray}
I&=& G(p)(e^A/\pi)\sum_{n=1}^{\infty }(-1)^{n+1}\frac{\sqrt{\pi }%
~\Gamma (np)}{2\Gamma (n)\Gamma \left(n(p-1)+3/2\right)}
\nonumber\\
&& \times\left( e^{A+\Delta
-\theta _{B}}\right) ^{2n(p-1)}\;,  \label{sbsi}
\end{eqnarray}
while the boundary condition (\ref{bc}) reads 
\begin{eqnarray}
U&=&2G(p) e^A-(p-1)G(p)e^A\sum_{n=1}^{\infty }(-1)^{n+1}
\nonumber\\
&&\times \frac{%
\sqrt{\pi }~\Gamma (np)}{\Gamma (n)\Gamma
\left(n(p-1)+3/2\right)}
\left(
e^{A+\Delta -\theta_{B}}\right)^{2n(p-1)} \;. \label{bdri}
\end{eqnarray}
Here we have introduced the notations 
$G(p)=p\sqrt{\pi} \Gamma(p/2(p-1)) \,/\,
\Gamma(1/2(p-1))$ 
and $\Delta=[\ln (p-1) -p\ln(p) /(p-1)]/2$.
The parameter $A$ follows by
solving Eq.~(\ref{bdri}) for a given 
voltage $U$.  Inserting it into Eq.~(\ref{sbsi})
gives the $I-U$ relation
and $W= 2 G(p) e^A$. Equations in the weak BS limit follow
from the duality relation described below.

The TBA equations are easily solved for any $T$ at $g=1/2$, with the result
\cite{egger98,foot}
\begin{equation} \label{g12}
V(W) = 2 T_B \,{\rm Im}\, \psi\left(\frac12
+ \frac{T_B + iW/2}{2\pi T} \right) \;,
\end{equation}
where $\psi$ is the digamma function and $T_B=\pi \lambda^2/\omega_c$.
The linear conductance is then
\begin{equation}
\label{lin}
G(T) = \frac{e^2}{h}
\,\frac{1-c\psi'(\frac12+c)}{1+c\psi'(\frac12+c)}\;,
\qquad c\equiv T_B/2\pi T \;,
\end{equation}
with the trigamma function $\psi'(x)$.
For high temperatures, this approaches $e^2/h$, while it
vanishes as $G\sim T^2$ 
at low temperatures.  Notably, while this is the same power
law as in the FQH effect \cite{kane92}, the prefactor is now different.
To get the current for arbitrary values of $g,T$ and $U$, 
one has to resort to a
straightforward numerical solution of the TBA equations (\ref{sbsi})
and (\ref{bdri}).
The linear conductance can be given in closed form,
\begin{equation}
\label{linearcond}
G=\frac{e^2}{h}\frac{\int \,d\theta \left| T_{++}\right|^{2}
df_\pm/d\theta } {\int \, d\theta \left( \left|
T_{++}\right| ^{2}+\left| T_{+-}\right| ^{2}/g \right) 
df_\pm/d\theta} \;,
\end{equation}
where the filling fractions (\ref{distr}) have to be evaluated at
$W=0$. 

A remarkable nonperturbative consequence of the TBA equations is the
existence of a duality relation for the current valid
at any temperature,
\begin{equation} \label{dual}
I(\lambda,U,g)={U\over 2\pi} -
 I\left(\lambda_d, U,{1\over g}\right) \;,
\end{equation}
where $\lambda_d\propto \lambda^{-1/g}$ \cite{saleurlate}. 
This duality is similar but different 
from  the one found
in the FQH case \cite{bcft1}. Although, like in the FQH case,
quasiparticles of charge $q=ge$ tunnel in the weak BS limit, and
electrons with $q=e$ in the strong BS limit, the coupling of the
QW to the reservoirs modifies the current and  gives rise to
the new duality (\ref{dual}). 

{\sl Bistability regime.}~--- The curves for the linear conductance 
in a QW are generally similar to those for the FQH
effect, with the main difference that the high temperature value is
now $e^2/h$ independent of $g$. Much more interesting and unexpected
physics occurs in the nonlinear out-of-equilibrium regime
for  very strong interactions and low temperatures.
The most direct approach to see this is the quasiclassical
limit, $g\ll 1$, at $T=0$, which we consider first.
In the bosonized theory, after integrating out
the standard boson phase fields away from $x=0$ but taking into
account the boundary conditions (\ref{bc}) \cite{egger98}, one is left
with the equation of motion 
\begin{equation}\label{eqm}
d\Phi/dt + g \pi T_B \sin\Phi = g W \;,
\end{equation}
where $T_B=2\lambda$ and the current operator is $\dot{\Phi}/2\pi$.
This equation can also be obtained from the explicit solution
 of the TBA in that limit. 
 For $W>\pi T_B$, 
the current grows proportional to
$\Delta= [W^2 - (\pi T_B)^2]^{1/2}$.  This is readily seen
from the  solution of Eq.~(\ref{eqm}), 
which reads in terms of $y=\tan[g\Delta t/2]$, 
\[
\tan(\Phi/2) = W y(t)/[\Delta + \pi T_B y(t)]
\;.
\]
Hence the average over one time period yields
\begin{equation} 
\langle I \rangle =\frac{g}{2\pi} \Delta\; \Theta(W-\pi T_B) \;,
\end{equation}
where $\Theta$ is the Heaviside function.
The $I(W)$ curve is single-valued, but by
inserting the definitions (\ref{defw}) and (\ref{defv}),
and eliminating $I$, we see that 
$U=F(W)$ with 
\begin{equation}
F(W) = gW + (1-g) V(W)  
\end{equation}
can have several solutions $W$.  Therefore the $I(U)$ relation
is multivalued throughout the regime $U_1<U<U_2$,
where $U_1=\sqrt{g(2-g)}\,\pi T_B$ and $U_2=\pi T_B$.  The three solutions
for the  current in this regime are $I=0$ and 
\begin{equation}
I_\pm (U) = \frac{U}{2\pi(2-g)}\left(1-g\pm \sqrt{1-g(2-g)(\pi T_B/U)^2}
\right) \;.
\end{equation}
For $U<U_1$, we only get $I=0$, and for $U>U_2$, $I_+(U)$ is the
only allowed solution.  Clearly, on the branch $I_-(U)$,
we have negative differential conductance (NDC), and therefore
this branch is unstable.  
Such S-shaped current-voltage relations are
familiar, e.g., in nonlinear semiconductor physics \cite{scholl}
and in charge-density wave transport \cite{cdw}.
By putting the QW into a properly designed load circuit,
self-sustained current oscillations can be generated \cite{scholl}.
Furthermore, putting a resistance $R$ in series and 
applying the voltage $V_a$ to the whole circuit, one has
\begin{equation}
V_a = RI(U) + U \;,
\end{equation}
which can easily be solved for $I(V_a)$.  For $R/(e^2/h)> g^{-1}-1$,
we get a single-valued $I(V_a)$ curve, i.e., the NDC branch has
been stabilized.  For smaller $R$, also the $I(V_a)$ curve
will be multi-valued.  In practice, one then gets hysteresis,
and the current jumps between two branches (bistability).

\begin{figure}
\epsfysize=6cm
\epsffile{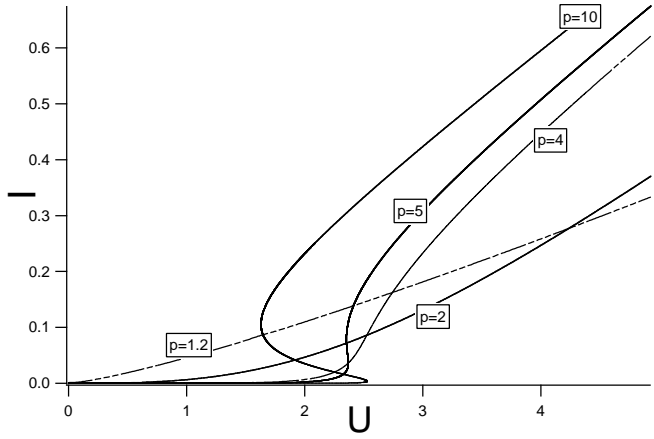}
\vspace{0.3cm}
\caption[]{\label{fig1}
Current-voltage relation at $T=0$ for various $g=1/p$.
The current is given in units of $(e^2/h)T_B$, and the
voltage in units of $T_B$.}
\end{figure}

How stable is this behavior once thermal and quantum
fluctuations are taken into account? 
Repeating the above $g\ll 1$ calculation 
for finite temperature with methods
from Ref.\cite{ivan}, one finds
\begin{equation}\label{high}
V(W) = \pi T_B  \,{\rm Im}\,
\frac{I_{1-iW/2\pi T}(T_B/2 T)}{I_{-iW/2\pi T}(T_B/2 T)}
\end{equation}
with the modified Bessel function $I_\nu(z)$ with complex order $\nu$.
The high-temperature expansion ($T\gg T_B$) of Eq.~(\ref{high}) predicts
bistability to occur for all $T<T_c(g)$ with 
\begin{equation}
T_c(g\ll 1) = T_B \sqrt{(1-g)/16g} \;.
\end{equation}
Therefore as $g\to 0$, the bistability occurs at all temperatures.
Bistability is not destroyed by quantum fluctuations either, as can 
be checked using the exact solution. At $T=0$, one finds that the NDC
disappears for $g\approx 0.2$, see Figure \ref{fig1}.
This value gets smoothly lowered as
temperature increases.
We stress that the bistability is a true
nonperturbative nonequilibrium phenomenon that has
no parallel in the linear conductance  \cite{footno}.

Finally we turn to  rather elementary physical reasoning
explaining the origin of the predicted bistable behavior
independent of the details of our boundary condition.
In the presence of impurity BS, a portion $V$ of the voltage
drops at the impurity site.  Then the average current 
$\langle I\rangle=(U-V)/2\pi$, where $1/2\pi$ [$=e^2/h$ in 
physical units] is the d.c.~conductance of the perfect wire.
However, a fluctuation $\delta I$ of the current must obey
\begin{equation}\label{di}
\delta I = -(g/2\pi) \delta V\;,
\end{equation}
with the associated fluctuation $\delta V$ 
of the four-terminal voltage and the a.c.~conductance
$g/2\pi$ \cite{kane92}. Equation~(\ref{di}) reflects
the fact that the impurity BS potential 
$\lambda \cos \Phi$ for the $x=0$ boson field $\Phi$
is due to tunneling of fractional quasiparticles with charge $ge$
\cite{kane94}. 
In terms of the boson field,
the current is $\langle I\rangle + \delta I = \dot{\Phi}/2\pi$,
and the voltage fluctuations read $\delta V = 2\pi \lambda \sin\Phi - V$.
The first term is basically the derivative of the pinning potential
$\lambda \cos\Phi$ \cite{kane94}.  With these relations, it
is straightforward to verify the equation of motion (\ref{eqm}).
The latter describes an overdamped particle moving in the tilted
washboard potential $-g(2\pi\lambda\cos\Phi + W\Phi)$, where the
bias $W = U+2\pi (g^{-1}-1) \langle I\rangle$ depends explicitly on the
current.  This feedback mechanism together with the
nonlinear pinning potential is responsible for bistability.
The above arguments also  suggest  that
our assumption of adiabatic coupling to the reservoirs is
not essential for bistability to occur \cite{foot4}. 

To conclude, nonequilibrium
transport through a spinless single-channel quantum wire
containing one impurity has been studied.
Using integrability techniques, the exact solution
of this interacting transport problem has been given for
adiabatically connected reservoirs.
We have discovered bistability phenomena in the
current for strong interactions
that should be observable in state-of-the-art experiments
and constitute a hallmark of strongly interacting quantum wires.

We thank L.~Balents, M.P.A.~Fisher, and F.~Guinea for discussions.
This work has been supported by the
Deutsche Forschungsgemeinschaft (Grant No.~GR 638/19-1),
by the National Science Foundation under Grants No.~PHY-94-07194 and
No.~PHY-93-57207, and by the DOE.

\begin{references}

\bibitem{book}
For a recent perspective, see A.O. Gogolin, A.A. Nersesyan, and
A.M. Tsvelik, {\sl Bosonization and Strongly Correlated
Systems} (Cambridge University Press, 1998).

\bibitem{tarucha95} S. Tarucha, T. Honda, and T. Saku,
Solid State Comm. {\bf 94}, 413 (1995).

\bibitem{yacoby96}
A. Yacoby {\em et al.}, Phys. Rev. Lett. {\bf 77}, 4612 (1996).

\bibitem{chain}
V. Vescoli {\sl et al.}, Science {\bf 281}, 1181 (1998).

\bibitem{tube} M. Bockrath {\em et al.},
Nature {\bf 397}, 598 (1999); Z. Yao {\em et al.}, {\em ibid.} 
{\bf 402}, 273 (1999).

\bibitem{chang}   A.M. Chang, L.N. Pfeiffer, and K.W. West,
Phys. Rev. Lett. {\bf 77}, 2538 (1996).

\bibitem{kane92} C.L. Kane and M.P.A. Fisher, Phys. Rev. B
{\bf 46}, 15 233 (1992).

\bibitem{bcft1}   P. Fendley, A.W.W. Ludwig, and H. Saleur,
Phys. Rev. B {\bf 52}, 8934 (1995);
P. Fendley and H. Saleur, Phys. Rev. B
{\bf 54}, 10 845 (1996).

\bibitem{scholl} E. Sch{\"o}ll, {\sl Nonequilibrium Phase Transitions
in Semiconductors} (Springer, Berlin, 1987).

\bibitem{landauer} R. Landauer, IBM J. Res. Dev.
{\bf 1}, 223 (1957); Z. Phys. B {\bf 68}, 217 (1987).

\bibitem{egger96} R. Egger and H. Grabert, Phys. Rev. Lett.
{\bf 77}, 538 (1996); {\em ibid.} {\bf 80}, 2255(E) (1998).

\bibitem{egger98} R. Egger and H. Grabert,
Phys. Rev. B {\bf 58}, 10 761 (1998).

\bibitem{saleurlate} 
A. Koutouza,  H. Saleur, and F. Siano, in preparation.

\bibitem{schulz}
D.L. Maslov and M. Stone, Phys. Rev. B {\bf 52}, R5539 (1995);
V.V. Ponomarenko, {\em ibid.} {\bf 52}, R8666 (1995);
I. Safi and H.J. Schulz, {\em ibid.} {\bf 52}, R17 040 (1995).

\bibitem{furu0}
D. Maslov, Phys. Rev. B {\bf 52}, R14 368 (1995);
A. Furusaki and N. Nagaosa, {\em ibid.} {\bf 54}, R5239 (1996).

\bibitem{safinew}
I. Safi, Eur. Phys. J B  {\bf 12}, 451 (1999).

\bibitem{buttiker}
Ya.M. Blanter, F.W.J.~Hekking, and M.~B{\"u}ttiker,
Phys. Rev. Lett. {\bf 81}, 1925 (1998).

\bibitem{foot}  Eq.~(5.30) in Ref.\onlinecite{egger98}
contains an incorrect numerical factor.
The corrected refermionization gives
indeed Eq.~(\ref{g12}).

\bibitem{cdw} G. Gr{\"u}ner, {\sl Density Waves in Solids}
(Addison-Wesley, Reading, 1994).

\bibitem{ivan} Yu.M. Ivanchenko and L.A. Zil'berman,
Zh Eksp. Teor. Fiz. {\bf 55}, 2395 (1968)
[Sov. Phys. JETP {\bf 28}, 1272 (1969)].

\bibitem{footno} 
Spin fluctuations will wash out the bistability. For an
experimental check, one has to study a spin-polarized QW.

\bibitem{kane94}
C.L. Kane and M.P.A. Fisher, Phys. Rev. Lett. {\bf 72}, 724 (1994).

\bibitem{foot4}
We expect that the relevant part of
the $T-g$ plane shrinks as the contact resistance increases.

\end{references}
\end{document}